\documentclass[conference]{IEEEtran}

\usepackage{cite,graphicx,amsmath,psfrag,bm}
\usepackage[usenames,dvipsnames]{xcolor}
\usepackage{mathabx}
\usepackage{subfig}
\usepackage{cancel}
\usepackage{epstopdf}
\usepackage{pstool}

\newcommand{\ve}[1]{\boldsymbol{#1}}
\newcommand{\te}[1]{\overline{\overline{#1}}}

\makeatletter\def\CT{\def\@captype{figure}}\makeatother

\begin{document}

\title{Synthesis of Electromagnetic Metasurfaces: Principles and Illustrations}

\author{\IEEEauthorblockN{Karim Achouri\IEEEauthorrefmark{1}, Bakthiar A. Khan\IEEEauthorrefmark{1}, Shulabh Gupta\IEEEauthorrefmark{1}, Guillaume Lavigne\IEEEauthorrefmark{1}, Mohamed A. Salem\IEEEauthorrefmark{1} and\\
Christophe Caloz\IEEEauthorrefmark{1}\IEEEauthorrefmark{2}}
\IEEEauthorblockA{\begin{minipage}{6.5cm}
\centering\linespread{1}
\IEEEauthorrefmark{1}Dept. of Electrical Engineering \\
Polytechnique Montr\'{e}al \\
Montr\'{e}al, QC H2T 1J3, Canada
\end{minipage}
\begin{minipage}{6.5cm}
\centering
\IEEEauthorrefmark{2}Electrical and Computer Engineering Dept. \\
King Abdulaziz University \\
POB 80204, Jeddah 21589, Saudi Arabia
\end{minipage}}}

\maketitle

\begin{abstract}
The paper presents partial overview of the mathematical synthesis and the physical realization of metasurfaces, and related illustrative examples. The synthesis consists in determining the exact tensorial surface susceptibility functions of the metasurface, based on generalized sheet transition conditions, while the realization deals with both metallic and dielectric scattering particle structures. The examples demonstrate the capabilities of the synthesis and realization techniques, thereby showing the plethora of possible metasurface field transmission and subsequent applications. The first example is the design of two diffraction engineering birefringent metasurfaces performing polarization beam splitting and orbital angular momentum multiplexing, respectively. Next, we discuss the concept of the ``transistor'' metasurface, which is an electromagnetic linear switch based on destructive interferences. Then, we introduce a non-reciprocal non-gyrotropic metasurface using a pick-up circuit radiator (PCR) architecture. Finally, the implementation of all-dielectric metasurfaces for spatial dispersion engineering is discussed.
\end{abstract}

\IEEEpeerreviewmaketitle


\section{Introduction}

Metasurfaces, which are the  two-dimensional counterparts of volume metamaterials~\cite{Holloway2009,holloway2012overview,yu2014flat}, have attracted much attention over the past years. Due to their low profile, small losses and rich electromagnetic field manipulation capabilities, they are excellent spatial processors able to manipulate electromagnetic waves with ever more complex possible applications, such as generalized refraction, polarization transformation, signal multiplexing, and non-reciprocal field control.

Metasurfaces are usually made of uniform or non-uniform arrangements of specifically engineered sub-wavelength scattering particles to produce a desired scattered field. Many such metasurfaces have been reported. However, these metasurface were most often designed using heavy optimization techniques. Fortunately, a few efficient synthesis techniques have been recently introduced, offering powerful and versatile tools for metasurface design~\cite{achouri2014general,Salem2013c,PhysRevApplied.2.044011,6477089,6905746}.

In this paper, we briefly recall the synthesis technique which is detailed in~\cite{achouri2014general}. Next, we illustrate the capabilities and the advantages of this synthesis technique. The various metasurfaces addressed here will be described and analysed more thoroughly in other future publications.

\section{Metasurface Design}

\subsection{Susceptibility Synthesis}
\label{sec:synthesis}

Let us consider that a metasurface lying in the $x-y$ plane at $z=0$. Given its inherent sub-wavelength thickness, a metasurface may be modeled as a zero-thickness spatial discontinuity. Rigorous boundary conditions, pertaining to such a discontinuity, were first derived by Idemen~\cite{Idemen1973} and later applied to metasurfaces by Kuester {\it et al.}~\cite{kuester2003av}. Conventionally called the Generalized Sheet Transitions Conditions (GSTCs), they read
\begin{subequations}
\label{eq:BCfinapp}
\begin{align}
\hat{z}\times\Delta\ve{H}
&=j\omega\ve{P}_{\parallel}-\hat{z}\times\nabla_{\parallel}M_{z},\label{eq:BCfinapp_1}\\
\Delta\ve{E}\times\hat{z}
&=j\omega\mu \ve{M}_{\parallel}-\nabla_{\parallel}\bigg(\frac{P_{z}}{\epsilon }\bigg)\times\hat{z},\label{eq:BCfinapp_2}\\
\hat{z}\cdot\Delta\ve{D}
&=-\nabla\cdot\ve{P}_{\parallel},\\
\hat{z}\cdot\Delta\ve{B}
&=-\mu \nabla\cdot\ve{M}_{\parallel},
\end{align}
\end{subequations}
where the time dependence $e^{j\omega t}$ is dropped by convenience. The permittivity $\epsilon$ and the permeability $\mu$ correspond to the medium surrounding the metasurface.  The sign~$\Delta$ corresponds to the difference of the electromagnetic fields $(\ve{E}, \ve{H}, \ve{D}$ and $\ve{B})$ between both sides of the metasurface, $\ve{P}$ and $\ve{M}$ are the electric and magnetic polarization densities, respectively.The subscript $\parallel$ refers to the vectorial components that are in the plane of the metasurface. The synthesis technique~\cite{achouri2014general} is based on the relations~\eqref{eq:BCfinapp} and determines the susceptibilities required to perform the specified wave transformations.

Relations~\eqref{eq:BCfinapp} form a set of coupled partial differential equations. To simplify the analysis, we consider that the metasurface is composed of only surface polarization densities, which reduces relations~\eqref{eq:BCfinapp_1} and~\eqref{eq:BCfinapp_2} to a set of coupled linear equations that can be solved for closed-form expressions of the metasurface susceptibilities. Upon substitution of the general definitions of $\ve{P}$ and $\ve{M}$, in terms of the bi-anisotropic susceptibility tensors $\te{\chi}_\text{ee}, \te{\chi}_\text{mm}, \te{\chi}_\text{em}$ and $\te{\chi}_\text{me}$ ~\cite{lindell1994electromagnetic},~\eqref{eq:BCfinapp} reduces to
\begin{subequations}
\label{eq:BC_plane}
\begin{align}
\hat{z}\times\Delta\ve{H}
&=j\omega\epsilon\te{\chi}_\text{ee}\ve{E}_\text{av}+jk\te{\chi}_\text{em}\ve{H}_\text{av},\label{eq:BC_plane_1}\\
\Delta\ve{E}\times\hat{z}
&=j\omega\mu \te{\chi}_\text{mm}\ve{H}_\text{av}+jk\te{\chi}_\text{me}\ve{E}_\text{av},\label{eq:BC_plane_2}
\end{align}
\end{subequations}
where the polarizations densities $\ve{P}$ and $\ve{M}$ are expressed in terms of the arithmetic average of the fields (as denoted by the subscript ``av") instead of the acting or total fields at the position of each scattering particles~\cite{kuester2003av}.

The mathematical synthesis of the metasurface essentially consists in inserting the electromagnetic fields of the specified transformations into~\eqref{eq:BC_plane} and solving for the susceptibilities. Depending on the desired transformation, the system of equations~\eqref{eq:BC_plane} may be under-determined, in which case some of the tensor elements must be set to zero for a definite solution~\cite{achouri2014general}. Once the susceptibilities of the metasurface are determined, the metasurface can be physically realized, which is the topic of the next section.

\subsection{Scattering Particle Synthesis and Implementation}
\label{sec:implem}

The physical realization of metasurfaces is, as of today, not a trivial routine. Here, two possible approaches are briefly discussed, while many more exist. In both approaches, the susceptibility functions, obtained by the aforementioned synthesis procedure, are discretized into unit cells. The unit cells are simulated (assuming periodic boundary conditions), one by one or group by group when the structure is fully or partly non-uniform, using commercial softwares that compute their scattering parameters. The required physical parameters for the scattering particles are obtained by mapping the scattering parameters into the susceptibility function~\cite{achouri2014general}. Finally, an enhanced design is achieved by tuning the parameters of the scattering particles via parametric analysis or standard optimization techniques.

\subsubsection{Metallic Scatterers}
\label{sec:implemCasc}

Metasurfaces consisting of metallic scatterers, typically on dielectric substrates, have been the most commonly reported types of metasurfaces, as well as their predecessors, the frequency selective surfaces (FSS)~\cite{MunkFSS}. In all cases, using more than one layer is an effective way to increase the available number of degrees of freedom, and hence achieving enhanced properties, including higher bandwidth and larger phase coverage range of the structure's unit cell. It was recently shown that three cascaded layers, where the two outer layers are identical, represents the minimum configuration to achieve full transmission and a $2\pi$ phase coverage~\cite{203903,6648706}. With three layers, the overall thickness of the metasurface generally remains sub-wavelength (usually in the order of $t\approx\lambda/10$) with negligible loss increase. Further increasing the number of layers may naturally be an approach for even broader bandwidth, at the expense of loss and weight.

\begin{figure}[h!]
\begin{center}
\includegraphics[width=0.5\columnwidth]{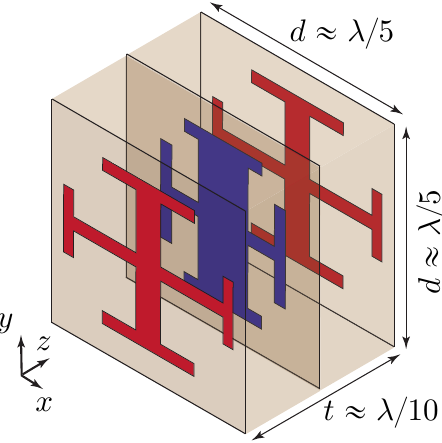}
\caption{Generic unit cell with three metallic (PEC) Jerusalem crosses separated by dielectric slabs, the outer layers are identical.}
\label{Fig:metallicUnitCell}
\end{center}
\end{figure}

A typical shape for the scattering particles forming the metallic layers is the Jerusalem cross, as shown in Fig.~\ref{Fig:metallicUnitCell}. The Jerusalem cross has the advantage of featuring fairly well decoupled responses for $x$ and $y$ polarizations, consequently simplifying the implementation. In the structure of Fig.~\ref{Fig:metallicUnitCell}, relatively strong capacitive coupling in the transverse $(x-y)$ plane offers the benefit of lowering the resonance frequencies~\cite{MunkFSS,hong2004microstrip} or, alternatively, of reducing the free-space electrical size of the unit cell ($d\approx\lambda/5$), while introducing more complexity in terms of coupling.

The realization of the unit cells can be greatly simplified if the longitudinal coupling between the three layers can be minimized, for example by decreasing the dielectric relative permittivity or/and increasing the dielectric thickness. In that case, each layer can be designed separately and the overall response of the multi-layer unit cell can then be found using simple transmission matrix approaches~\cite{6648706}.

\subsubsection{Dielectric Scatterers}
\label{sec:dielc}

\begin{figure}[h!]
\begin{center}
\subfloat[]{\label{Fig:Conventional_MS1}
\includegraphics[width=0.45\columnwidth]{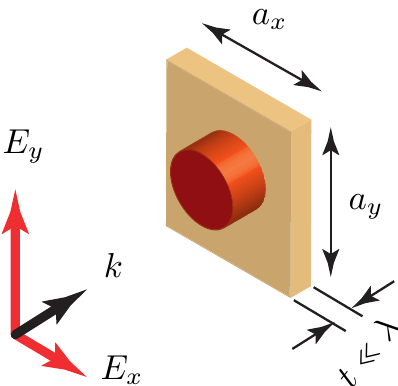}
}
\subfloat[]{\raisebox{10mm}{\label{Fig:Conventional_MS2}
\includegraphics[width=0.45\columnwidth]{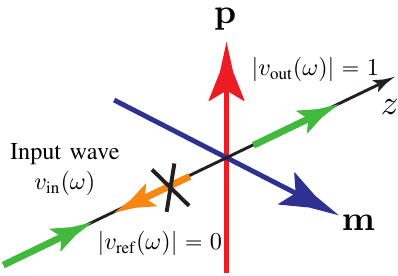}
}}
\caption{(a) Representation of an all-dielectric all-pass metasurface unit cell consisting of a dielectric resonator ($\varepsilon_{r,1}$) embedded in a host layer of permittivity $\varepsilon_{r,2}$. (b)~Operation principle for full transmission (zero reflection).}
\label{Fig:Conventional_MS}
\end{center}
\end{figure}

The second implementation method is based on all-dielectric scattering particles. It has been shown~\cite{kiv,arbabi2014complete} that dielectric resonators exhibit both electric and magnetic resonances. A typical unit cell is shown in Fig.~\ref{Fig:Conventional_MS1} where the particles are dielectric cylinders of circular cross section with permittivity $\varepsilon_{r,1}$ embedded in a substrate with permittivity $\varepsilon_{r,2}$. Other types of particle shapes are naturally also possible. It is possible, by tuning the physical dimensions of the resonator as well as the permittivities ratio $\varepsilon_{r,1}/\varepsilon_{r,2}$, to merge the electric and magnetic resonances. In this scenario, if the two resonances have the same strength and are associated with orthogonal dipole moments in the transverse plane of the metasurface, as shown in Fig.~\ref{Fig:Conventional_MS2}, reflection may be totally suppressed. This is due to perfect destructive interference of the waves scattered by the electric and magnetic scattering particles in the incident side of the metasurface, and their constructive interference at the transmission side of it. In this case, the transmission is theoretically $100\%$ and flat over a wide bandwidth. Moreover, the transmission phase covers a full $2\pi$ range around the resonance frequency $\omega$. This powerful concept may therefore apply to all-pass metasurfaces with controllable phase over large bandwidth.

Structurally symmetric shapes like cylinders or squares present the same behavior for $x$- and $y$-polarized waves. However, using asymmetric shapes, such as ellipses and rectangles, allow for a complete and independent control of the two orthogonal polarizations, as recently demonstrated in~\cite{arbabi2014complete}. As an additional advantage, dielectric unit cells have a greatly reduced number of physical parameters to adjust compared to the three layer Jerusalem crosses of Fig.~\ref{Fig:metallicUnitCell}, effectively simplifying the optimization procedure to achieve the specified response.

\section{Illustrative Examples}

\subsection{Birefringent Metasurface}

The first example we present is a birefringent metasurface. Birefringence is the property, available in certain crystal, according to which different refractive indices are presented to different wave polarizations. This property requires anisotropy~\cite{saleh2007fundamentals}. A birefringent crystal has, in the plane transverse to propagation, a fast axis and slow axis corresponding to low and high refractive indices, respectively. Thus, a wave polarized along the direction of the fast axis undergoes a smaller phase shift than a wave polarized along the slow axis. Based on this principle, several interesting devices can be realized. For example, a half-wave plate, which induces a $\pi$ phase shift between the two orthogonal polarizations so as to change the polarization angle of linearly polarized waves (rotation of polarization without chirality) or change the handedness of circularly polarized waves. Similarly, a quarter-wave plate induces a $\pi/2$ phase shift between the two polarizations, which allows transforming a linearly polarized wave into a circularly polarized wave or vice versa.

Birefringent metasurfaces can be mathematically described using the synthesis technique of Sec.~\ref{sec:synthesis}. Since bi-anisotropy (or chirality) are not required to achieve birefringence,~\eqref{eq:BC_plane} reduces to
\begin{subequations}
\label{eq:InvProb}
\begin{equation}
\label{eq:diffH}
\binom{-\Delta H_y}{\Delta H_x}=j\omega\epsilon  \begin{pmatrix} \chi_{\text{ee}}^{xx} & 0 \\ 0 & \chi_{\text{ee}}^{yy} \end{pmatrix}\binom{E_{x,\text{av}}}{E_{y,\text{av}}},
\end{equation}
\begin{equation}
\label{eq:diffE}
\binom{\Delta E_y}{-\Delta E_x}=j\omega\mu  \begin{pmatrix} \chi_{\text{mm}}^{xx} & 0 \\ 0 & \chi_{\text{mm}}^{yy} \end{pmatrix}\binom{H_{x,\text{av}}}{H_{y,\text{av}}}.
\end{equation}
\end{subequations}
Following the procedure described in~\cite{achouri2014general}, the system is straightforwardly solved and yields two sets of orthogonal susceptibilities, given for $x$-polarized waves by
\begin{subequations}
\label{eq:chi_diag_x}
\begin{align}
\chi_{\text{ee}}^{xx}&=\frac{-\Delta H_{y}}{j\omega\epsilon  E_{x,\text{av}}},\label{eq:chi_diag_Exx}\\
\chi_{\text{mm}}^{yy}&=\frac{-\Delta E_{x}}{j\omega\mu  H_{y,\text{av}}},\label{eq:chi_diag_Myy}
\end{align}
\end{subequations}
and for $y$-polarized waves by
\begin{subequations}
\label{eq:chi_diag_y}
\begin{align}
\chi_{\text{ee}}^{yy}&=\frac{\Delta H_{x}}{j\omega\epsilon  E_{y,\text{av}}},\label{eq:chi_diag_Eyy}\\
\chi_{\text{mm}}^{xx}&=\frac{\Delta E_{y}}{j\omega\mu  H_{x,\text{av}}}.\label{eq:chi_diag_Mxx}
\end{align}
\end{subequations}
We shall next consider two specific types of birefringences, generalized refraction and orbital angular momentum birefringences, that are both achieved by using proper non-uniform susceptibilities in~\eqref{eq:chi_diag_x} and~\eqref{eq:chi_diag_y}.

\subsubsection{Generalized Refraction Birefringence}
\label{sec:splitter}

\begin{figure}[h!]
\centering
\includegraphics[width=0.88\columnwidth]{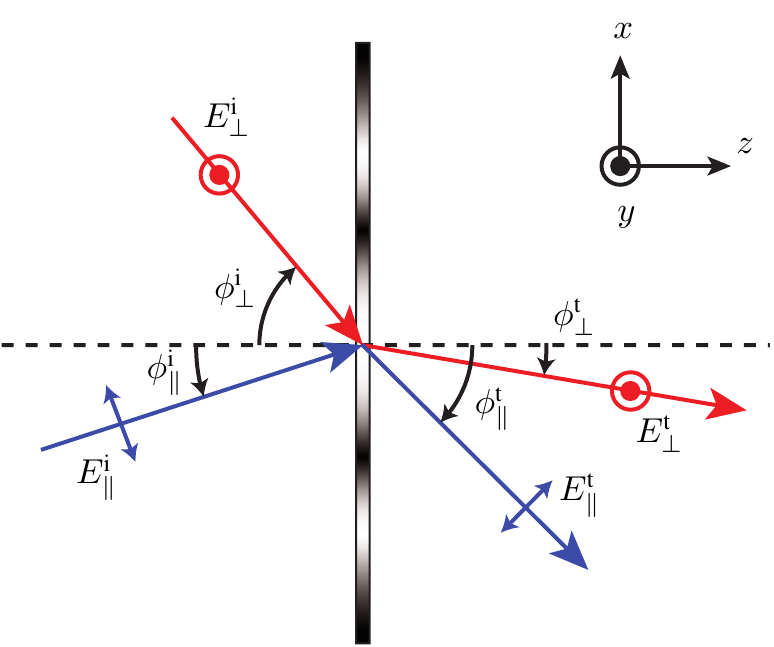}
\caption{Example of a birefringent ``generalized refraction''. Two orthogonally polarized plane waves incident on the metasurface get refracted at different and independent angles.}
\label{fig:schem}
\end{figure}

The concept of generalized refraction birefringence is proposed here as a direct application of the general synthesis technique described in~\cite{achouri2014general} and simplified for the case of birefringence in~\eqref{eq:chi_diag_x} and~\eqref{eq:chi_diag_y}. Generalized refraction birefringence consists in independently and simultaneously controlling the reflection and the transmission coefficients of $x$-polarized and $y$-polarized plane waves incident on the metasurface. The principle is illustrated in Fig~\ref{fig:schem}, where the reflected waves are specified to be zero.

\begin{figure}[h!]
\begin{center}
\subfloat[]{\label{Fig:polsplit1}
\includegraphics[width=0.6\columnwidth]{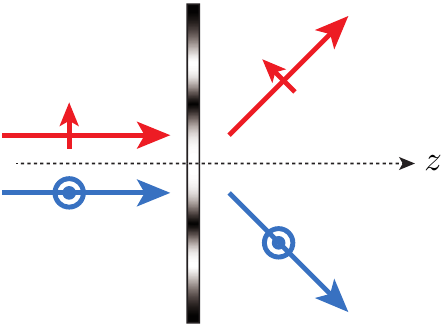}
}
\\
\subfloat[]{\label{Fig:polsplit2}
\includegraphics[width=0.75\columnwidth]{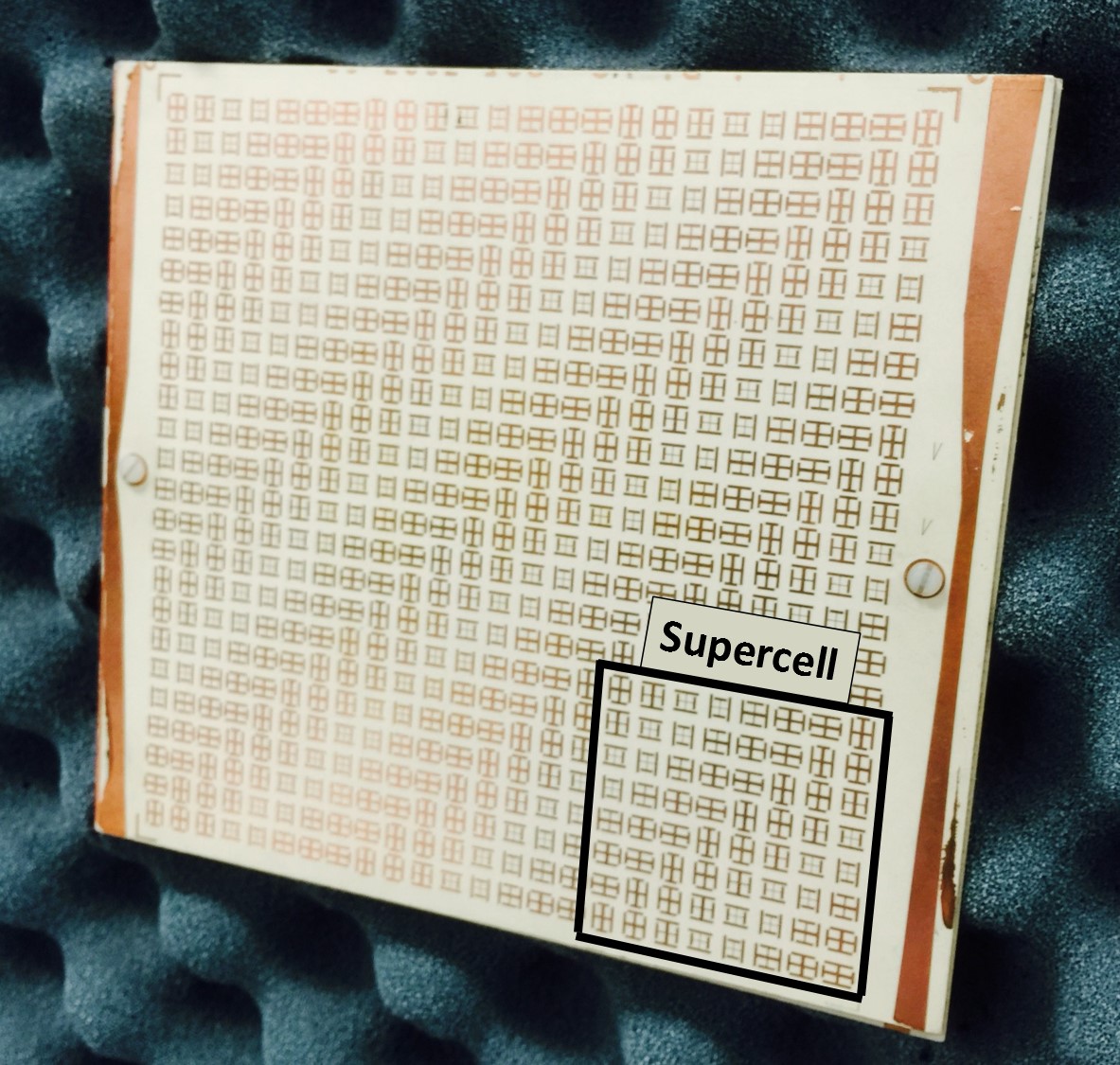}
}
\caption{(a) Representation of the polarization beam splitter/combiner. (b) Fabricated structure with $24\times24$ unit cells. The two polarizations are refracted in the diagonal direction where a macroscopic stripe-like pattern is apparent.}\label{Fig:polsplit}
\end{center}
\end{figure}

As an example, we present the implementation of a polarization beam splitter (or, reciprocally, a polarization combiner). The corresponding metasurface, that is shown in Fig.~\ref{Fig:polsplit}, is first described in terms of its surface susceptibilities, by inserting the specified incident, reflected and transmitted electric and magnetic fields into~\eqref{eq:chi_diag_x} and~\eqref{eq:chi_diag_y}. Here, the reflected waves are specified to be zero and the transmitted waves are  specified to be fully refracted at $\phi_\perp=+60^\circ$ and $\phi_\parallel=-60^\circ$, respectively, in the diagonal of the square metasurface. The computed susceptibilities, which are periodic in the $x-y$ plane, are then discretized into $8$ fundamental unit cells of sides $d\approx \lambda/5$ and thickness $t\approx \lambda/10$. Following the procedure described in Sec.~\ref{sec:implemCasc}, each unit cell is designed separately to realize the required equivalent surface susceptibilities. The fabricated metasurface, shown in Fig.~\ref{Fig:polsplit2}, consists in $24\times 24$ unit cells. From periodicity, the unit cells can be grouped into periodic supercells composed of $8\times 8$ unit cells.
%

\begin{figure}[h!]
\begin{center}
\subfloat[]{\label{Fig:SimPolSplitt1}
\includegraphics[width=1\columnwidth]{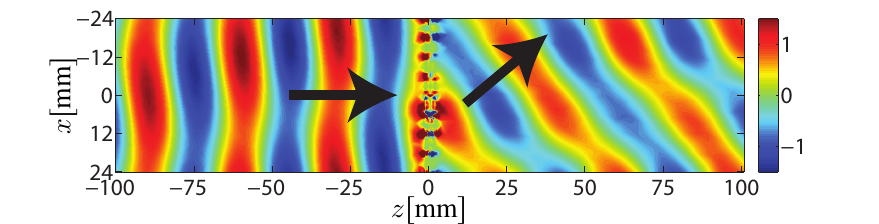}
}
\\
\subfloat[]{\label{Fig:SimPolSplitt2}
\includegraphics[width=1\columnwidth]{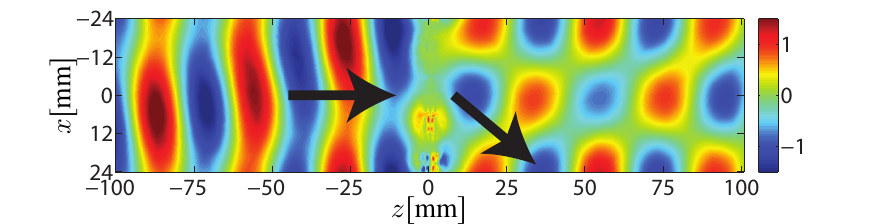}
}
\caption{Full-wave simulations, assuming periodic boundary conditions, of the first row of the supercell shown in Fig.~\ref{Fig:polsplit2}. (a) An $x$-polarized normally incident plane wave is impinging from left to right on the metasurface that refracts it upward. (b) A $y$-polarized normally incident plane wave is impinging from left to right on the metasurface that refracts it downward.}\label{Fig:SimPolSplitt}
\end{center}
\end{figure}

Because the metasurface in Fig.~\ref{Fig:polsplit2} is electrically too large to be simulated, a simplified version of the metasurface is considered for the full-wave simulations. Only the first row of the $8\times 8$ supercell is selected. Numerical simulations (CST) are performed for $x$ and $y$ polarizations. The results are shown in Fig.~\ref{Fig:SimPolSplitt1} and Fig.~\ref{Fig:SimPolSplitt2}, respectively. It has to be noted that, because the periodicity is now only in the $x$ direction, instead of being in the diagonal of the supercell, the refraction angle is reduced. This can be understood by considering that the projected phase of the refracted wave on the metasurface undergoes one full cycle along the $x$ direction but two full cycles along the diagonal of the supercell.

As may be seen in Fig.~\ref{Fig:SimPolSplitt1}, the simulation is in good agreement with the expected result. However, the simulation in Fig.~\ref{Fig:SimPolSplitt2} shows a refracted wave perturbed by the presence of a zeroth diffraction order contribution. The reason for the mediocre results of Fig.~\ref{Fig:SimPolSplitt2} is the altered coupling between the unit cells due to truncation, which yields the observed zeroth order transmission.

\subsubsection{Orbital Angular Momentum Birefringence}

The second example for a birefringent metasurface is an orbital angular momentum (OAM), or vortex wave, multiplexing metasurface~\cite{nye1974dislocations}. The generation of OAM waves using metasurfaces has been presented in several recent reports~\cite{karimi2014generating,Yi:14,PhysRevApplied.2.044012}. Bessel waves are a typical choice for vortex waves. However, Bessel waves are either radially or azimuthally polarized, demanding for either inhomogeneous rotation of polarization if the incident plane wave is linearly polarized, or that the incident plane wave be circularly polarized. To avoid this constraint, Hypergeometric-Gaussian (HyG) waves are considered instead. The electric field of a HyG wave is given by~\cite{karimiHYG}
\begin{equation}
\begin{split}
\label{eq:HYG}
E(\rho,\phi,z) =& \frac{\Gamma \left ( 1+|m|+\frac{p}{2}  \right )}{\Gamma (|m|+1)}\frac{i^{|m|+1}\zeta^{|m|/2}\xi^{p/2}}{[\xi+i]^{1+|m|/2+p/2}}e^{im\phi-i\zeta}\\
&\times _{1}F_1 \left ( -\frac{p}{2},|m|+1;\frac{\zeta[\xi + i]}{\xi[\xi - i]} \right ),
\end{split}
\end{equation}
where $_{1}F_1(a,b;x)$ is the confluent hypergeometric function, $\Gamma(x)$ is the gamma function, $m$ is the OAM order, $p\geqslant -|m|$ is a real parameter, and where $\zeta = \rho^2/(w_0^2[\xi+i])$, $\xi=z/z_R$ $w_0$, with $w_0$ being the beam waist and $z_r$ the Rayleigh range given by $z_r=\pi w_0^2/\lambda$.

Compared to Bessel waves, HyG waves have the advantage of being linearly polarized, thus allowing direct transformation (i.e. no rotation of polarization) from a linearly polarized plane wave, as well as the multiplexing of orthogonal plane waves into orthogonal vortex HyG of different order $m$.

\begin{figure}[h!]
\begin{center}
\subfloat[]{\label{Fig:HYG1}
\includegraphics[width=0.5\columnwidth]{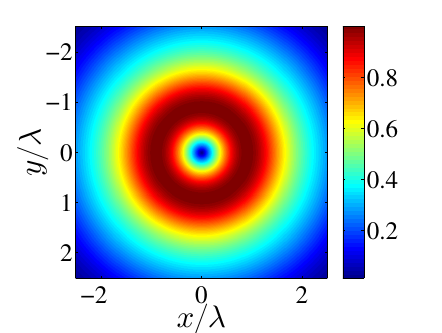}
}
\subfloat[]{\label{Fig:HYG2}
\includegraphics[width=0.5\columnwidth]{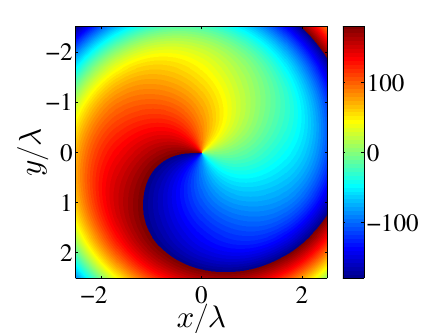}
}
\\
\subfloat[]{\label{Fig:HYG3}
\includegraphics[width=0.5\columnwidth]{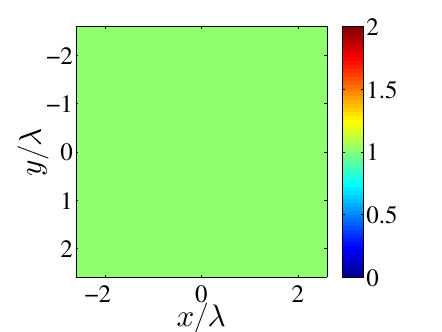}
}
\subfloat[]{\label{Fig:HYG4}
\includegraphics[width=0.5\columnwidth]{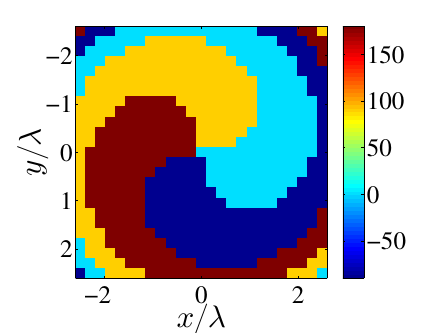}
}
\caption{ Magnitude (a) and phase (b) of the Hypergeometric-Gaussian (HyG) wave with parameters $p=1$, $m=-1$, $w_0=\lambda$, $\xi=1 $. Simplified version of the specified fields with flat magnitude (c) and $4$ phase samples (d).}\label{Fig:HYG}
\end{center}
\end{figure}

Consider the following transformation, where two normally incident plane waves, polarized along $x$ and $y$, respectively, are transformed into HyG waves of order $m=+1$ and $m=-1$, respectively. Figure~\ref{Fig:HYG} shows the typical amplitude and phase distributions of a vortex HyG wave of order $m=-1$. Because of its complex structure, a complete discretization of the HyG wave would result in a very large number of scattering particles. In order to simplify the problem, the HyG wave phase is coarsely sampled with only $4$ phase samples, as shown in Fig.~\ref{Fig:HYG4}. The $4$ phase samples are sufficient, according to the Nyquist criterion, to describe the full phase evolution of the wave. As for the magnitude of the HyG wave, it is transformed into a flat uniform plane of magnitude $|E|=1$, as shown in Fig.~\ref{Fig:HYG3}. Although the implemented magnitude is flat, the radiated field magnitude variations follow naturally from self-interference.

\begin{figure}[h!]
\begin{center}
\subfloat[]{\label{Fig:HYGsim1}
\includegraphics[width=0.5\columnwidth]{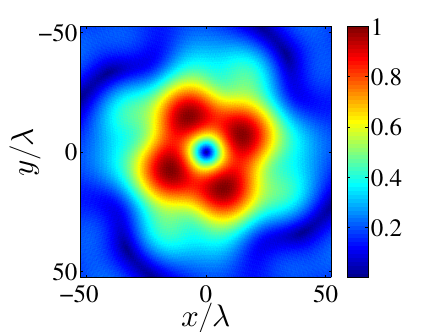}
}
\subfloat[]{\label{Fig:HYGsim2}
\includegraphics[width=0.5\columnwidth]{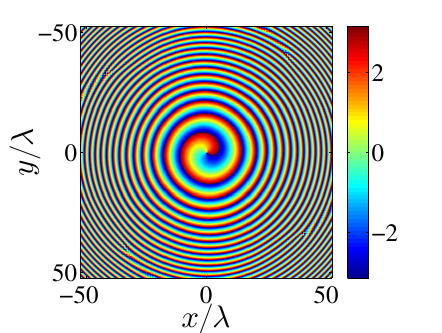}
}
\\
\subfloat[]{\label{Fig:HYGsim1}
\includegraphics[width=0.5\columnwidth]{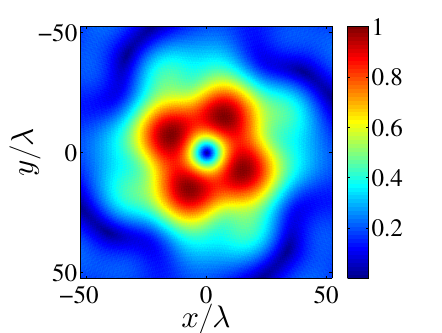}
}
\subfloat[]{\label{Fig:HYGsim2}
\includegraphics[width=0.5\columnwidth]{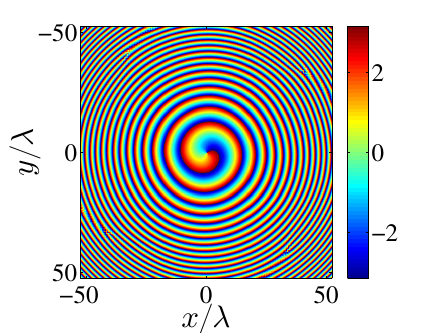}
}
\caption{Fourier propagated (a) magnitude and (b) phase, at a distance of $z = 100\lambda$, generated by the simplified fields of Fig.~\ref{Fig:HYG3} and Fig.~\ref{Fig:HYG4}. Fourier propagated (c) magnitude and (d) phase, at a distance of $z = 100\lambda$, for the HyG wave of order $m=+1$. The $4$ stronger regions in (a) and (c) are due to the square shape of the aperture formed by the metasurface $(\text{size } 5\lambda\times5\lambda)$. Using a circular metasurface would result in a better approximation of the ideal doughnut shaped magnitude of Fig.~\ref{Fig:HYG1}.}\label{Fig:HYGsim}
\end{center}
\end{figure}

It can be easily shown that, even with such a drastic simplification, the fields scattered by the metasurface do exhibit the smooth shape of the continuous fields of Fig.~\ref{Fig:HYG1} and Fig.~\ref{Fig:HYG2}. This can be verified by Fourier propagating the fields of Fig.~\ref{Fig:HYG3} and Fig.~\ref{Fig:HYG4}, as shown in Fig.~\ref{Fig:HYGsim}. Similar considerations apply to the HyG wave of order $m=+1$ whose phase spirals in the opposite direction than the phase of the HyG wave of order $m=-1$.

\begin{figure}[h!]
\begin{center}
\includegraphics[width=0.75\columnwidth]{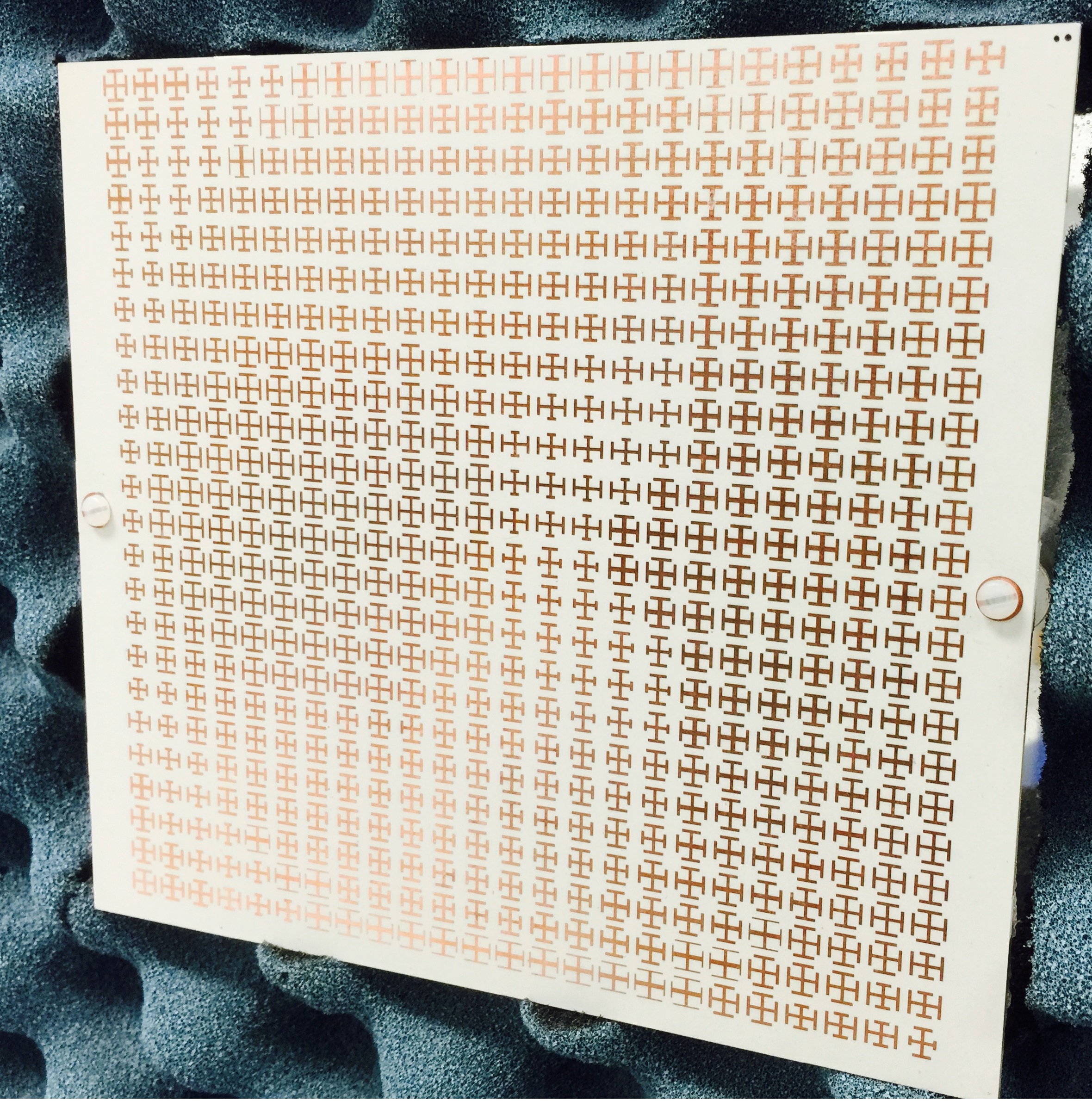}
\caption{Fabricated OAM multiplexing metasurface.}\label{Fig:HYGstruc}
\end{center}
\end{figure}

The final metasurface is implemented by combining both $m=+1$ and $m=-1$ HyG waves, which results in a total number of $16$ different unit cells distributed over the area of the metasurface. The final fabricated structure is shown in Fig.~\ref{Fig:HYGstruc}.

\subsection{Transistor Metasurface}

In this section, we propose to synthesize and implement a ``transistor'' metasurface whose transmitted wave can be modified by the presence of a control incident wave in a similar fashion as the output signal of a transistor is modulated by an external signal. The metasurface has two main operation states, with or without control wave. In the most general case, all waves can be arbitrarily specified and the control wave can therefore dramatically modify the initial transmitted wave. Moreover, changing the amplitude, phase or polarization of the control wave separately can be used as additional degrees of freedom to tune the transmitted wave. The metasurface itself is not necessarily tunable, in contrast to certain structures with tunable scattering particles~\cite{Burokur2,PhysRevB.86.195408,6349393,zhu2013active}. Likewise, the electromagnetic properties of the metasurface are not modified by the presence of the control wave in contrast to, for instance~\cite{xie2013spatial,harris2008electromagnetically}.

\begin{figure}[h!]
\centering
\subfloat[]{\includegraphics[width=0.4\linewidth]{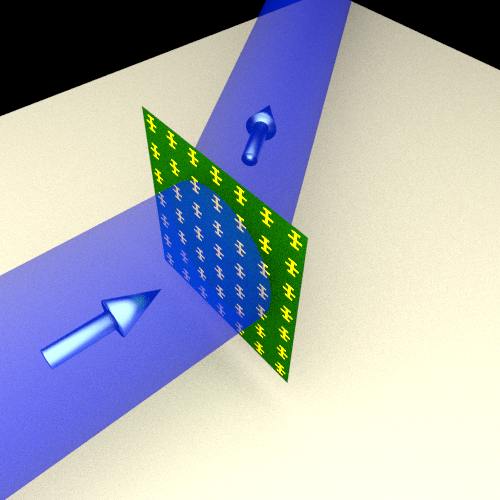}
\label{fig:param_prob1}}
\qquad
\subfloat[]{\includegraphics[width=0.4\linewidth]{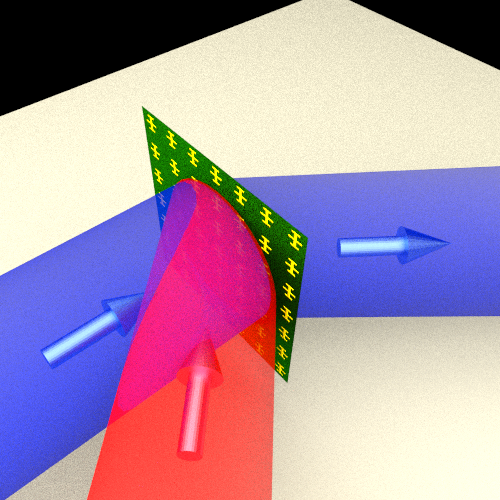}
\label{fig:param_prob2}}
\caption{Transistor metasurface concept. The metasurface is represented as an array of Jerusalem crosses. In this example, zero reflection is specified. (a) An incident wave (or signal) is transformed into a transmitted wave refracted at $45^\circ$. (b) When a control wave (in red) is impinging on the metasurface, the transmitted wave refraction angle becomes $-45^\circ$. Note that both signal and control waves have the same frequency.}\label{fig:param_prob}
\end{figure}

Consider for instance the transistor metasurface illustrated in Fig.~\ref{fig:param_prob}. In this example, the reflected waves are specified to be zero. When no control wave is present, the metasurface transforms an incident wave (or signal) into a transmitted wave refracted at a $45^\circ$ angle with respect to the normal of the metasurface, as shown in Fig.~\ref{fig:param_prob1}. When the control wave is impinging on the metasurface, as in Fig.~\ref{fig:param_prob2}, the transmitted wave refraction angle becomes $-45^\circ$.

\begin{figure}[h!]
\begin{center}
\subfloat[]{\label{Fig:TMconcept1}
\includegraphics[width=0.5\columnwidth]{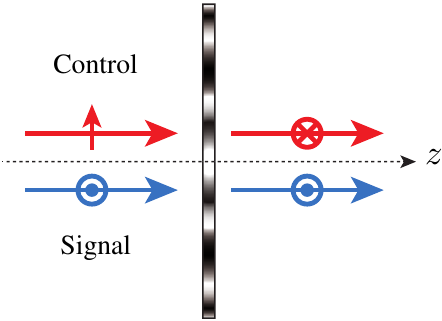}
}
\subfloat[]{\label{Fig:TMconcept2}
\includegraphics[width=0.5\columnwidth]{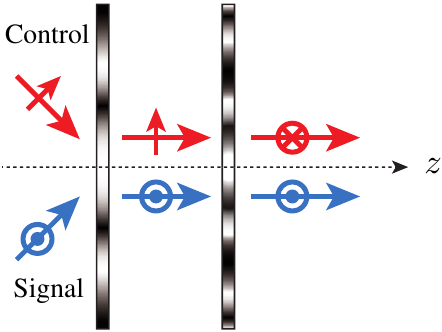}
}
\caption{Representations of the switching transistor metasurface concept. (a) The control and the signal waves are both normally incident on the metasurface. The polarization of the control wave is rotated to match the polarization of the signal wave. A $\pi$-phase shift is imposed between the two transmitted waves, so that they cancel each other by destructive interference. (b) The control and the signal waves are first combined together by the polarization combiner described in Sec.~\ref{sec:splitter}.}\label{Fig:TMconcept}
\end{center}
\end{figure}

We now present a simpler operation for the transistor metasurface concept: switching. Consider two plane waves, one $x$-polarized (the signal wave) and one $y$-polarized (the control wave), both normally incident on the transistor metasurface. The signal wave is normally transmitted by the metasurface, whereas the control wave is also normally transmitted but its polarization is rotated to match the polarization of the transmitted signal wave. The control is achieved by destructive interference between the signal and the control waves, as shown in Fig~\ref{Fig:TMconcept1}. In the configuration of Fig~\ref{Fig:TMconcept1}, the signal wave and the control wave have the same point of incidence and their respective sources would therefore overlap. To overcome this difficulty, we propose the configuration of Fig~\ref{Fig:TMconcept2}, where the metasurface presented in Sec.~\ref{sec:splitter} is used as a polarization combiner instead of a polarization beam splitter.
The physical implementation of the transistor metasurface presented in Fig.~\ref{Fig:TMconcept} inherently requires chirality and could therefore not be directly realized with the concept of birefringence. To synthesize the metasurface of Fig.~\ref{Fig:TMconcept1}, the system~\eqref{eq:BC_plane} can be solved by specifying appropriate scattering parameters, as discussed in~\cite{achouri2014general,PhysRevApplied.2.044011}. One possible set of scattering parameters is given here
%
\begin{subequations}
\label{eq:Sparam}
\begin{equation}
\label{eq:S21}
\te{S_{21}}= \frac{\sqrt{2}}{2}\begin{pmatrix} 0 & 0 \\ -1 & 1 \end{pmatrix},
\end{equation}
\begin{equation}
\label{eq:S12}
\te{S_{12}}= \frac{\sqrt{2}}{2}\begin{pmatrix} 0 & -1 \\ 0 & 1 \end{pmatrix},
\end{equation}
\end{subequations}
where port $1$ corresponds to the incident side of Fig.~\ref{Fig:TMconcept1} (left-hand side) and port $2$ corresponds to the transmit side (right-hand side). The reflection scattering tensors $\te{S_{11}}$ and $\te{S_{22}}$ are here left as free parameters. Due to the chiral nature of the metasurface, the structure is bi-anisotropic irrespectively to the values of $\te{S_{11}}$ and $\te{S_{22}}$. An important point, apparent in relations~\eqref{eq:Sparam}, is that the efficiency of the transistor metasurface described here is limited to $50\%$. This can be understood by considering a $y$-polarized normally incident plane wave from port $2$. As given by~\eqref{eq:S12}, the $y$-polarized incident wave splits equally into two waves respectively polarized along $x$ and $y$. Since the metasurface is reciprocal (i.e. $\te{S_{21}} = \te{S_{12}}^T$), the transmission of either the signal or the control wave is limited to $50\%$. Higher efficiency could be achieved with a non-reciprocal metasurface which is the topic of the next section.

\begin{figure}[h!]
\begin{center}
\subfloat[]{\label{Fig:TMstruc1}
\includegraphics[width=0.5\columnwidth]{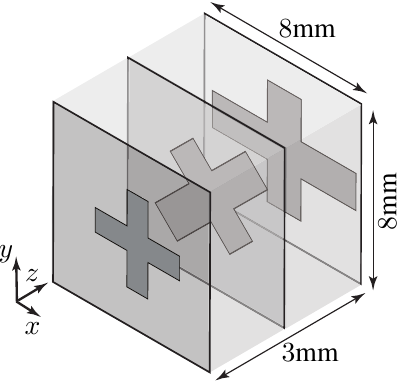}
}
\\
\subfloat[]{\label{Fig:TMstruc2}
\includegraphics[width=0.6\columnwidth]{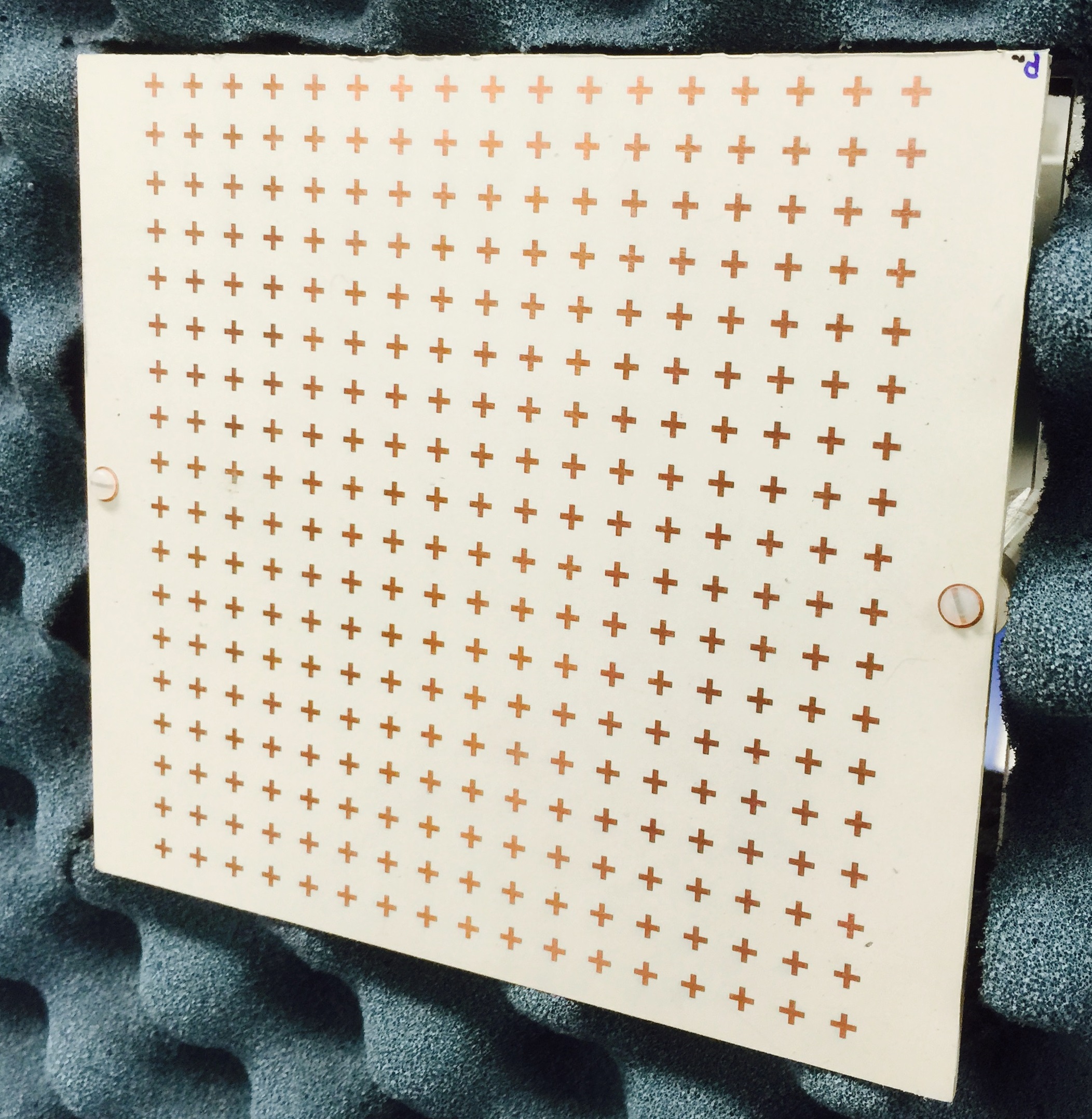}
}
\caption{(a) Representation of a three layer unit cell of the transistor metasurface. (b) Fabricated structure, with $17\times18$ unit cells, on the measurement stage.}\label{Fig:TMstruc}
\end{center}
\end{figure}

To implement the switching transistor metasurface, the three metallic layer approach is used. To obtain the chiral specified behavior, the central metallic layer is rotated $45^\circ$, as shown in Fig.~\ref{Fig:TMstruc}.

\begin{figure}[h!]
\centering
\CT
\subfloat[]{\label{Fig:C}
\includegraphics[width=0.9\columnwidth]{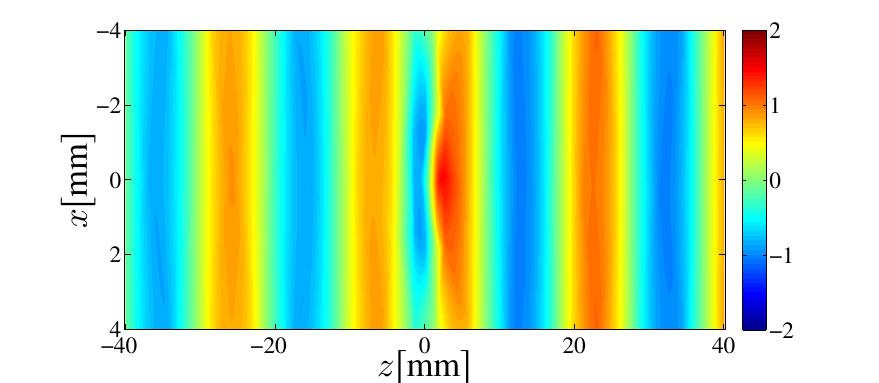}
}
\\
\subfloat[]{\label{Fig:S}
\includegraphics[width=0.9\columnwidth]{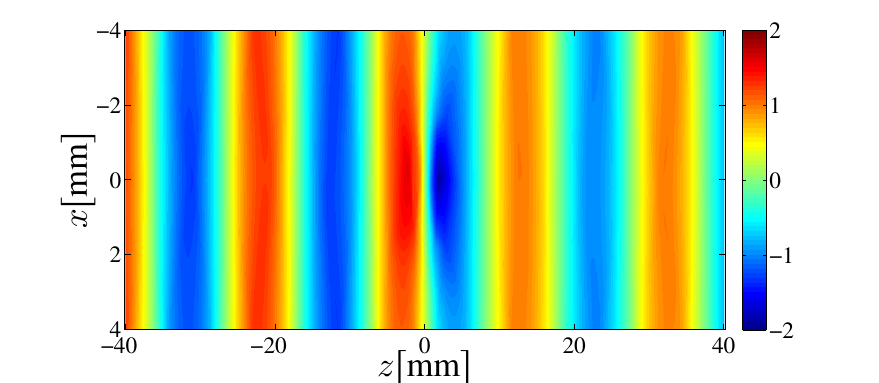}
}
\\
\subfloat[]{\label{Fig:CS}
\includegraphics[width=0.9\columnwidth]{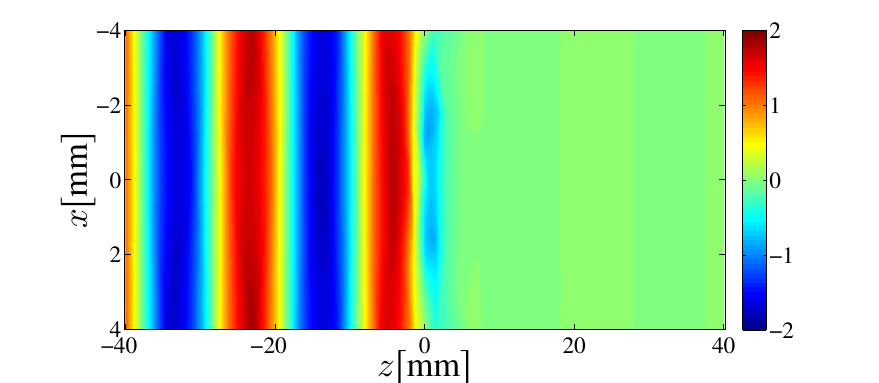}
}
\caption{Numerical simulations (CST) of the $y-$polarized fields, the metasurface is at $z=0$. (a) Fields when only the control wave is incident on the structure. The control wave is initially $x$-polarized (not shown here) from left to right. In the positive $z$-direction, the control wave is transmitted with a $y$ polarization. In the negative $z$-direction, the control wave is reflected with a $y$ polarization. (b) Fields when only the signal wave is incident on the metasurface. (c) Combination of control and signal waves, the cancellation of the transmission by destructive interference is clearly visible.}\label{Fig:TMsim}
\end{figure}

Simulations of the metasurface when illuminated by the signal wave, the control wave and with the two waves simultaneously is shown in Fig.~\ref{Fig:C},~\ref{Fig:S} and~\ref{Fig:CS}, respectively.

\subsection{Non-Reciprocal Non-Gyrotropic Metasurface}

Non-reciprocal devices are usually based on magnetic material non-reciprocal gyrotropy, whose Faraday rotation is a particular effect~\cite{2,3}. Recently, non-reciprocal gyrotropic metasurfaces, having a similar response to magnetic materials but without requiring a magnet, have been reported~\cite{kodera2009non,kodera2009nonradome,parsa2011ferrite}.

\begin{figure}[h!]
\begin{center}
\subfloat[]{\label{Fig:NRNGa}
\includegraphics[width=0.9\columnwidth]{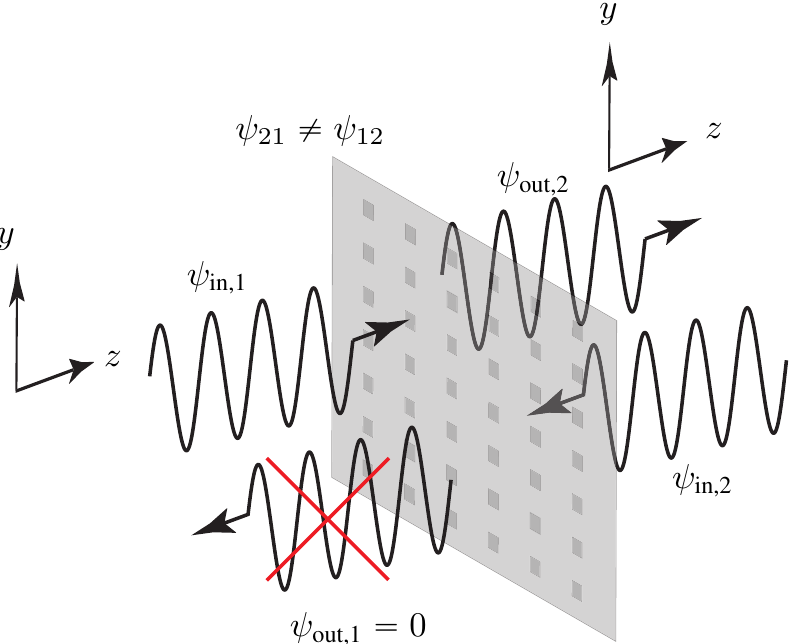}
}
\\
\subfloat[]{\label{Fig:NRNGb}
\includegraphics[width=0.9\columnwidth]{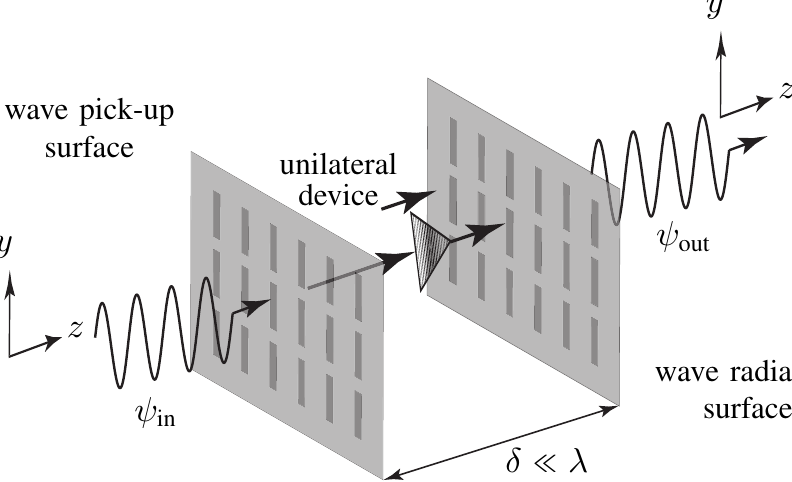}
}
\caption{Spatial nonreciprocal nongyrotropic metasurface. a)~ Functional illustration. b)~ Proposed pick up-circuit-radiator (PCR) implementation.}\label{Fig:NRNG}
\end{center}
\end{figure}

We will discuss here a one-wave transparent metasurface that does not induce polarization rotation and whose principle is shown in Fig.~\ref{Fig:NRNGa}.
The specified response of the structure is
\label{eq:Sparam2}
\begin{equation}
\label{eq:S21}
\te{S_{21}}= \begin{pmatrix} 1 & 0 \\ 0 & 1 \end{pmatrix},
\end{equation}
while the other scattering parameters are $\te{S_{12}}=\te{S_{11}}=\te{S_{22}}=0$. Solving the system~\eqref{eq:BC_plane}, following the procedure described in~\cite{achouri2014general,PhysRevApplied.2.044011}, with the specified scattering parameters, yields the susceptibility tensors
\begin{subequations}
\label{eq:ChiTrans}
\begin{equation}
\label{eq:ChiTrans1}
\te{\chi}_\text{ee}= -\frac{j}{k}\begin{pmatrix} 1 & 0 \\ 0 & 1 \end{pmatrix},
\quad
\te{\chi}_\text{mm}= -\frac{j}{k}\begin{pmatrix} 1 & 0 \\ 0 & 1 \end{pmatrix},
\end{equation}
\begin{equation}
\label{eq:ChiTrans2}
\te{\chi}_\text{em}= \frac{j}{k}\begin{pmatrix} 0 & 1 \\ -1 & 0 \end{pmatrix},
\quad
\te{\chi}_\text{me}= \frac{j}{k}\begin{pmatrix} 0 & -1 \\ 1 & 0 \end{pmatrix}.
\end{equation}
\end{subequations}
The non-reciprocal behavior of the metasurface is evident from the fact that the susceptibility tensors~\eqref{eq:ChiTrans2} dot not respect the reciprocity condition, i.e. violate $\te{\chi}_\text{em} = -\te{\chi}_\text{me}^T$. It can be easily understood why the synthesized metasurface needs to be bi-anisotropic even though no rotation of polarization is specified: because it is specified to be non-gyrotropic, the non-diagonal elements of $\te{\chi}_\text{ee}$ and $\te{\chi}_\text{mm}$, as well as the diagonal elements of $\te{\chi}_\text{em}$ and $\te{\chi}_\text{me}$, are necessarily zero since they correspond to gyrotropic behavior. Consequently, the tensors $\te{\chi}_\text{ee}$ and $\te{\chi}_\text{mm}$, being diagonal, can only correspond to reciprocal behavior according to the reciprocity conditions (i.e. $\te{\chi}_\text{ee} = \te{\chi}_\text{ee}^T$ and $\te{\chi}_\text{mm} = \te{\chi}_\text{mm}^T$). The remaining way to simultaneously achieve non-reciprocity and non-gyrotropy in such case is by leveraging bi-anisotropy.

\begin{figure}[h!]
\begin{center}
\subfloat[]{
\includegraphics[width=0.6\columnwidth]{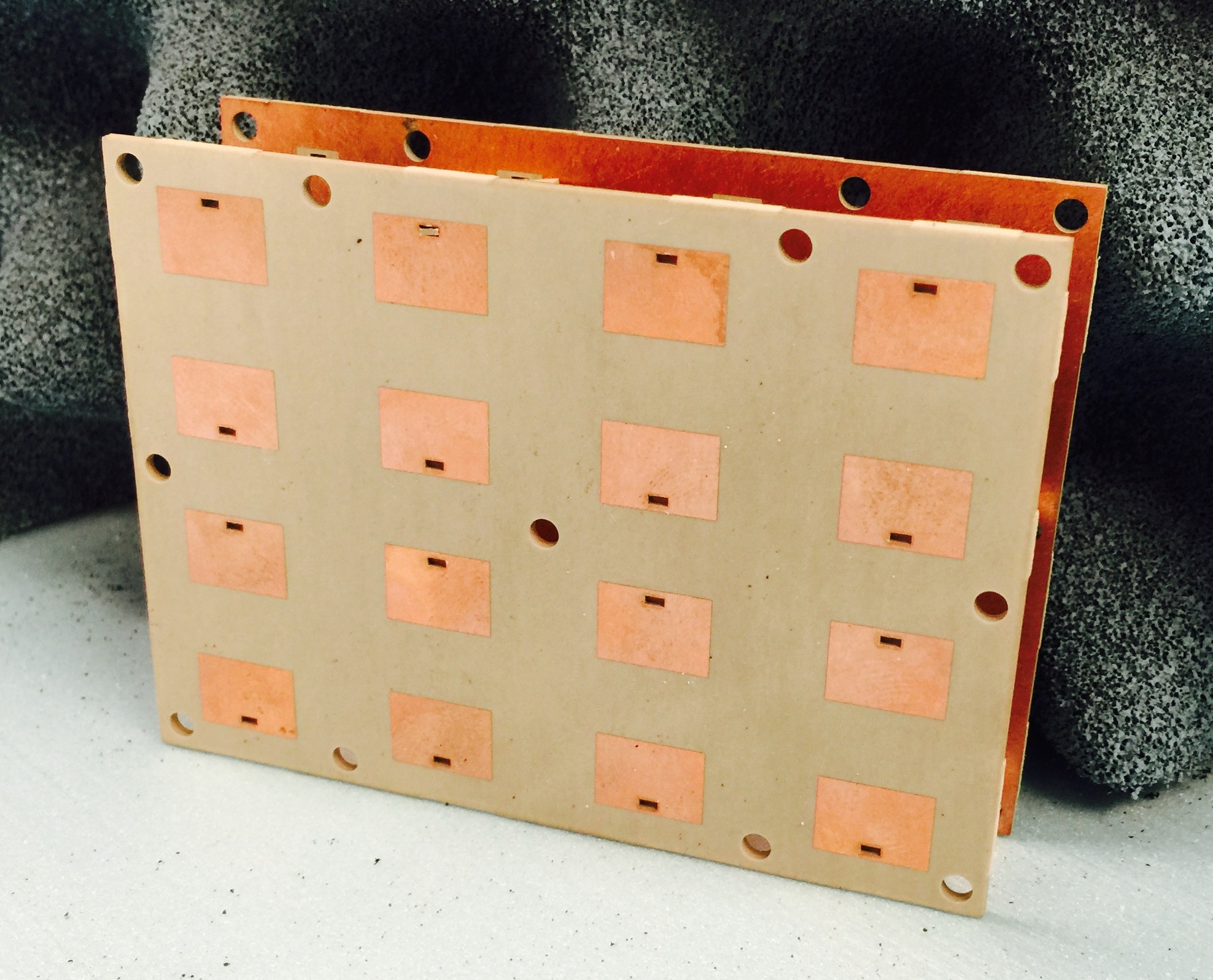}}
\\
\subfloat[]{
\includegraphics[width=0.48\columnwidth]{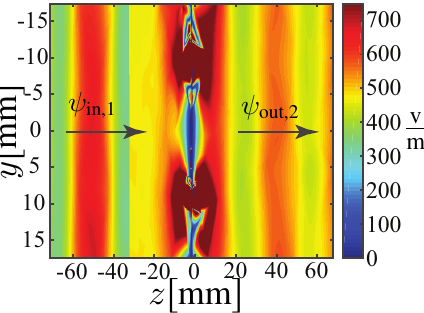}
}
\subfloat[]{
\includegraphics[width=0.48\columnwidth]{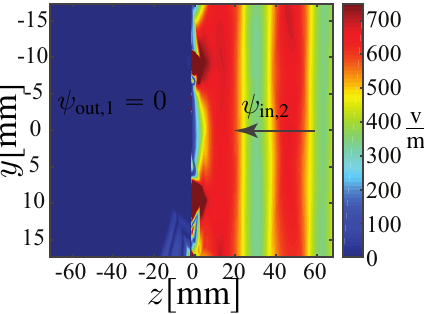}
}
\caption{(a) Fabricated non-reciprocal non-gyrotropic metasurface with $16$ unit cells.  Full-wave simulated electric fields of the metasurface loaded with a perfect isolator when: (b) excited from Port 1 and (c)~excited from Port 2. The operating frequency is $9.5$ GHz.}\label{Fig:NRNGsim}
\end{center}
\end{figure}

Because the structure described by relations~\eqref{eq:ChiTrans} is not trivial to implement with the tools given in Sec.~\ref{sec:implem}, we have used an alternative approach based on a similar design to the one used in~\cite{wang2012gyrotropic}. The metasurface, as shown in Fig.~\ref{Fig:NRNGb}, consists in a pick-up circuit radiator (PCR) structure. The pick-up and radiator faces of the metasurface are composed of patch antennas linked together by an electric circuit present in the center of the structure. When the electric circuit is loaded with an insulating device (like a FET transistor), transmission is allowed only in one direction, effectively realizing the specified scattering parameters. More details about the implementation of this metasurface will be presented elsewhere. Figures~\ref{Fig:NRNGsim} show the simulated field when the metasurface is illuminated from left to right or from right to left, the non-reciprocal behavior can clearly be seen.

The realization of non-reciprocal metasurfaces may become an important part in the implementation of more complicated metamaterial structures allowing more complex field transformations as, for example, the transistor metasurface discussed in the previous section.

\subsection{Spatial Dispersion Engineering}

So far, most existing metasurfaces perform spatial transformations for monochromatic waves only. As metasurface applications will become more and more advanced, the demand for non-uniform phase control over large frequency bands will increase. As it stands, the synthesis technique of Sec~\ref{sec:synthesis} is restricted to monochromatic wave transformations. However,~\eqref{eq:BC_plane} may be solved repeatedly for different frequencies, consequently forming frequency-dependent, or temporally-dispersive, susceptibility components, i.e. $\chi(\omega)$. The response of the metasurface could therefore be engineered for the specified frequency band in the similar fashion that phasers can achieve real-time analog signal processing (R-ASP)~\cite{Caloz_MM_2012}.

\begin{figure}[h!]
\centering
\includegraphics[width=0.8\columnwidth]{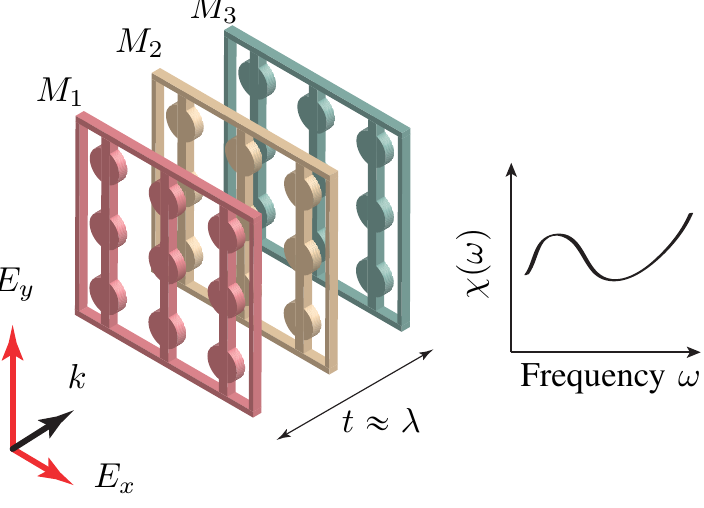}
\caption{Conceptual idea of cascading several dielectric metasurfaces for dispersion engineering.}\label{Fig:Cascade}
\end{figure}

\begin{figure}[h!]
\begin{center}
\subfloat[]{
\includegraphics[width=0.6\columnwidth]{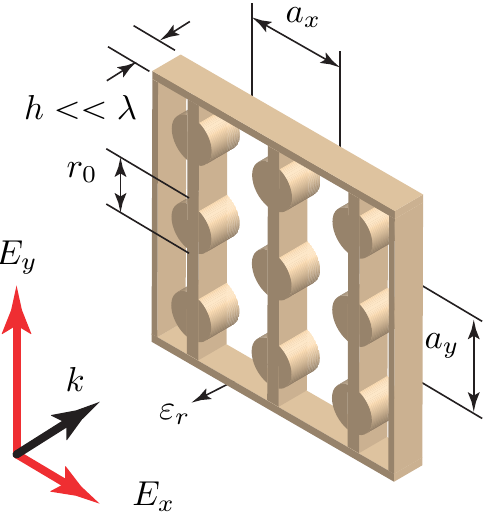}
}
\\
\subfloat[]{\label{Fig:Proposed_MS2}
\includegraphics[width=0.8\columnwidth]{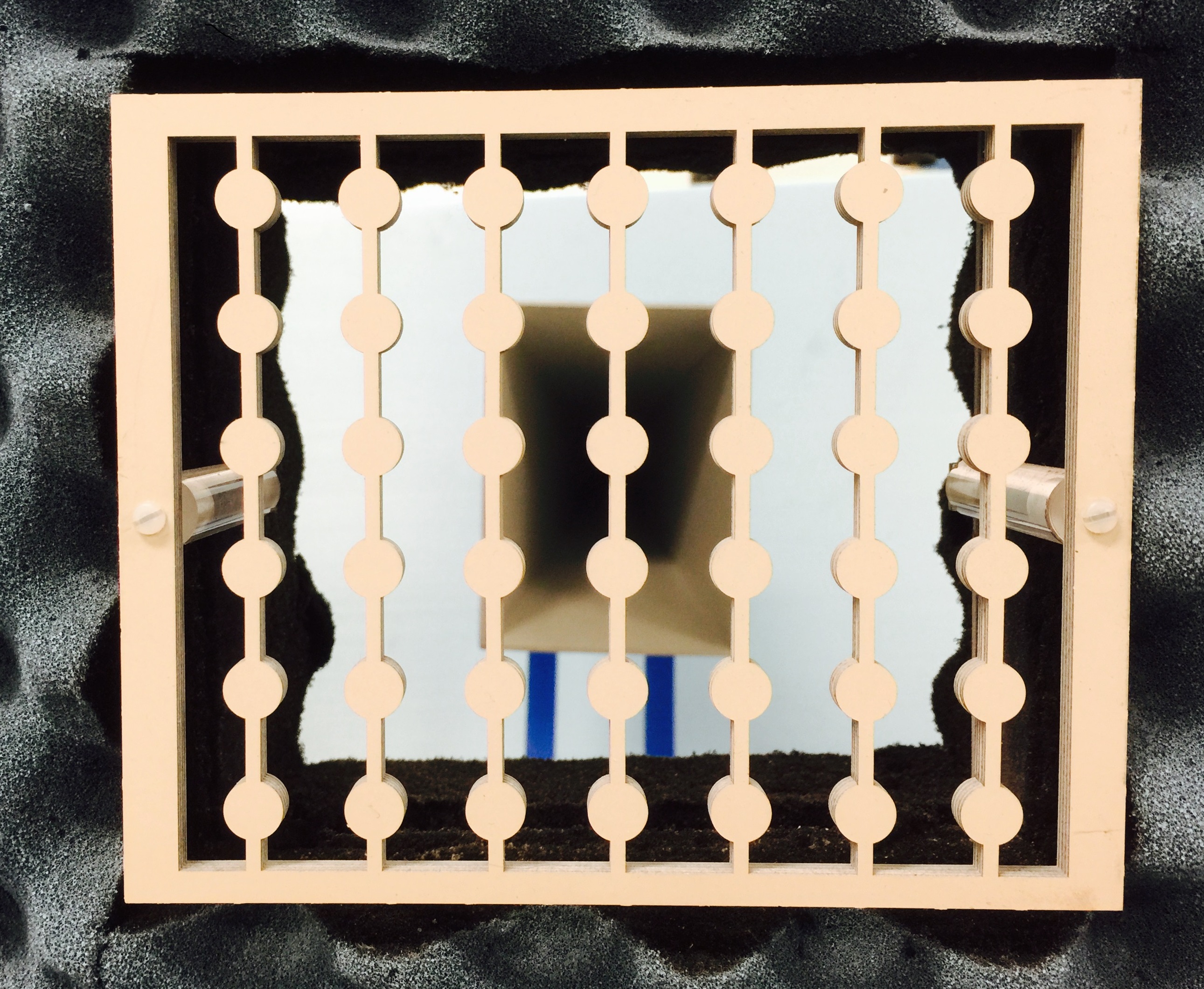}}
\caption{Proposed all-pass and all-dielectric metasurface based on 2-D array of inter-connected coupled cylinders. (a) Structure illustration. (b) Fabricated prototype, the exciting horn antenna is visible in the background.}
\label{Fig:Proposed_MS}
\end{center}
\end{figure}

One possible approach, to engineer frequency-dependent metasurfaces is to use the additional degrees of freedom provided by coupled resonators as demonstrated in~\cite{Aieta20032015}. Another possibility, that is proposed here, is to cascade metasurfaces made of dielectric scattering particles, as depicted in Fig.~\ref{Fig:Cascade}. As shown in Sec.~\ref{sec:dielc}, cylindrical dielectric resonators can be tuned to achieve full transmission and $2\pi$ phase coverage over a specific bandwidth. The proposed all-pass metasurface is shown in Fig.~\ref{Fig:Proposed_MS}. In this configuration, each cylinder is mechanically supported by dielectric interconnections, which also suppress spurious Fabry-P\'{e}rot resonances that would occur if the dielectric particles were placed on a supporting slab.

\begin{figure}[h!]
\centering
\label{Fig:DielecSim1}
\includegraphics[width=1\columnwidth]{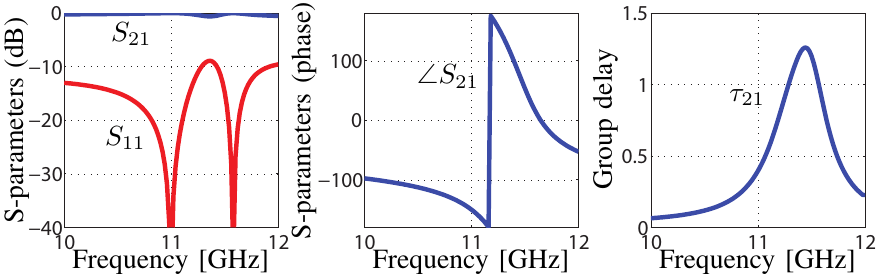}
\caption{Rigorous coupled wave analysis (RCWA) of the metasurface of Fig.~\ref{Fig:Proposed_MS2}. The plots show the transmission and reflection coefficients as well as the transmission phase and the transmission groupe delay, respectively. The field is polarized in the $x$ direction (perpendicular to the dielectric interconnections).}\label{Fig:DielecSim}
\end{figure}

The simulated response of the dielectric metasurface of Fig.~\ref{Fig:Proposed_MS2} is presented in Fig.~\ref{Fig:DielecSim}. A flat and almost full transmission is achieved over a large bandwidth while a full phase cycle occurs around the resonant frequency $\omega = 11.15$~GHz.

\section{Conclusion}

In this paper, we have presented a partial overview of the synthesis and possible applications of electromagnetic metasurfaces and several illustrative examples of metasurface applications, in an increasing order of synthesis complexity. First, the concept of birefringence, which allows an independent control of both orthogonal polarizations, was used to implement two diffraction engineering metasurfaces: one for polarization dependent generalized refraction and the other one for orbital angular momentum multiplexing. Next, the implementation of a transistor metasurface was discussed. It turns out that birefringence, which is based on anisotropy, is fundamentally insufficient to realize this transistor metasurface, the latter requiring bi-anisotropic constituents. It was shown that the transistor metasurface efficiency is limited due to its inherent reciprocal operating principle. However, a non-reciprocal transistor metasurface could theoretically achieve a much better efficiency. Based on this consideration, a non-reciprocal non-gyrotropic metasurface was introduced. This metasurface, which also requires bi-anisotropic features, could be a possible future solution to realize efficient non-reciprocal transistor metasurfaces. Finally, we have introduced a conceptual idea to realize all-pass dispersion engineering metasurfaces able to control the group delay of electromagnetic signals of large bandwidth, which constitute a missing piece in the metasurface arsenal.

\bibliographystyle{IEEEtran}
\bibliography{LIB}

\end{document}